\documentclass[pra,preprint,showpacs,showkeys,superscriptaddress]{revtex4}
\usepackage{hyperref}
\usepackage{amsfonts}
\usepackage{amsmath}
\usepackage{amssymb}
\usepackage{graphicx}

\begin{document}

\title{Experimental Simulation of a Pairing Hamiltonian on an NMR Quantum
Computer}
\author{Xiao-Dong Yang}
\email{xdyang77@ustc.edu.cn}
\affiliation{Department of Modern Physics, University of Science and Technology of China,
Hefei 230026, People's Republic of China}
\author{An Min Wang}
\email{anmwang@ustc.edu.cn}
\affiliation{Department of Modern Physics, University of Science and Technology of China,
Hefei 230026, People's Republic of China}
\author{Feng Xu}
\email{xxufeng@mail.ustc.edu.cn}
\affiliation{Department of Modern Physics, University of Science and Technology of China,
Hefei 230026, People's Republic of China}
\author{Jiangfeng Du}
\email{djf@ustc.edu.cn}
\affiliation{Department of Modern Physics, University of Science and Technology of China,
Hefei 230026, People's Republic of China}
\affiliation{Hefei National Laboratory for Physical Sciences at Microscale, Hefei 230026,
People's Republic of China}
\affiliation{Department of Physics, National University of Singapore, Lower Kent Ridge,\\
Singapore 119260, Singapore}
\date{\today }

\begin{abstract}
We have developed a concrete quantum simulation scheme and experimentally
simulated a pairing model on an NMR quantum computer. The design of our
experiment includes choosing an appropriate initial state in order to make
our scheme scalable in near future, and the accomplishment of our experiment
makes use of twice Fourier transforms so that our method is applicable to
other physical models. Our results show that the experimental simulation can
give the spectrum of the simulated Hamiltonian. Consequently, the potential
power of a quantum computer on the simulation of complex physical systems is
verified.
\end{abstract}

\maketitle

\section{Introduction}

Quantum computer (QC) can offer attractive ability in accelerating the
computation. The most appealing feature from the classical computer, as
Feynman \cite{feyn} noted, is that QC can be used to simulate the physical
behavior of a real system. Lloyd \cite{lloyd} later confirmed this idea in a
two-state array. Recently, Somaroo \textit{et al.} \cite{cory1} presented a
general scheme for the quantum simulation with a physical system to simulate
another. At present, among the physical realizations of QC, nuclear magnetic
resonance (NMR) has shown the greatest achievements (for a review see e.g.,
\cite{cory2,jones}) though it uses the nuclear spin ensemble \cite%
{cory3,gershenfeld}. With NMR technique, some interesting physical issues
have been simulated successfully, including a four-level truncated quantum
harmonic oscillator \cite{cory1}, a natural decoherence of a two-spin system
\cite{tseng1}, a two-qubit correlation function of the Fano-Anderson model
\cite{ortiz}, a three-spin effective Hamiltonian \cite{tseng2}, and the
migration of excitation in an eight-state quantum system \cite{fung}.

Quantum simulation utilizes a physical system, commonly referred to a QC,
which is easier to control and measure, to simulate the dynamics of another
one, which is rather a complex system, also called the simulated system.
Quantum simulation is expected to solve the problems in quantum many-body
theory which is dealt with more difficult by the classical methods. The
simulated system we will discuss is the BCS model \cite{bcs,mahan} which is
of wide interest in condensed matter and gives a phenomenal explanation to
superconductivity. In BCS model, one of the key problem is to find the
energy gap, \textit{i.e.}, the energy required at least to excite a
quasiparticle from the Fermi surface \cite{philip}. In fact, some
approximate methods \cite{duk,rich} have been proposed to calculate the
energy gap for a given BCS Hamiltonian. However, if we realize the quantum
simulation, one can get the accurate result within a polynomial time and
resources \cite{lawu}.

In this paper, we show that QC can be used to simulate the property of a
superconductor, such as the BCS model. Based on a concrete scheme of quantum
simulation on NMR QC which includes a method for choosing an appropriate
working initial state and the usage of twice Fourier transforms \cite{wang2}%
, we experimentally obtained the spectrum of the BCS Hamiltonian, which
coincided with the theoretic expectations. In Ref. \cite{sommapra}, authors
had mentioned that the energy spectrum can be obtained by a fast Fourier
transform after the measurement. In this paper, we deepen this point and
show how to implement the measurement of the spectrum on an NMR QC.

\section{Scheme of quantum simulation on an NMR QC}

For a complete quantum simulation procedure, there are three mainly
qualitative steps -- Hamiltonian mapping, the experimental simulation, and
again mapping back to the simulated system. We will obey these three steps
to describe our quantum simulation scheme of the BCS model in the following.

\textit{Firstly, the mapping from the BCS Hamiltonian to the Pauli
operators. }

The reduced Hamiltonian of the BCS model is a pairing Hamiltonian \cite%
{philip,lawu, xu},

\begin{equation}
H_{BCS}=\hbar \lbrack \sum_{m=1}^{N}\frac{\varepsilon _{m}}{2}%
(n_{m}+n_{-m})+V\sum_{m,l=1}^{N}c_{m}^{\dag }c_{-m}^{\dag }c_{-l}c_{l}],
\label{bcs}
\end{equation}%
where $\varepsilon _{m}$ denotes the free electron energy from the Fermi
surface; $V$ the coupling coefficient where it is simplified as a constant
\cite{ash}; $n_{\pm m}\equiv c_{\pm m}^{\dagger }c_{\pm m}$ the electron
number operators; $c_{m}^{+}(c_{m})$ the fermionic creation (annihilation)
operator; $m=1,2,\cdots ,N$ represent all of relevant quantum numbers, and
the electron pairs are labelled by the the quantum number $m$ and $-m$,
according to the Cooper pair situation where the paired electrons have equal
energies but opposite momenta and spins: $m=(\overrightarrow{k},\uparrow )$
and $-m=(-\overrightarrow{k},\downarrow )$. For a typical metal
superconductor, $\hbar \varepsilon _{m}\sim 10^{-2}$ eV, $\hbar V\sim
10^{-6} $ eV \cite{xu}.

Our interested physical property of the BCS superconductor is the spectrum
of the Hamiltonian (\ref{bcs}), because an important parameter in
superconductor --- energy gap can be followed from it \cite{taylor}. The
energy gap in superconductor is the energy difference between the ground
state of the element excitation to its first excited state. A lot of
analytical and numerical methods (see \cite{rom} for a review) have been
developed to obtain the energy gap through the diagonalization of the BCS
Hamiltonian. However, these methods are complicated and use some unproved
approximations. In this paper, we show that the quantum simulation on a QC
can solve it in a practical and direct way with reduced resources.

In Ref. \cite{philip,lawu}, authors had\textit{\ }mapped the Hamiltonian (%
\ref{bcs}) into the qubit space based on the isomorphic algebras of
spin-fermion connection. The mapped Hamiltonian is also known as the
spin-analogy,

\begin{equation}
H_{p}=\hbar \lbrack \sum_{m=1}^{N}\frac{\varepsilon _{m}}{2}\sigma _{z}^{m}+%
\frac{V}{2}\sum_{l>m=1}^{N}(\sigma _{x}^{m}\sigma _{x}^{l}+\sigma
_{y}^{m}\sigma _{y}^{l})],  \label{p}
\end{equation}%
where $\sigma _{x},\sigma _{y},\sigma _{z}$ denote the Pauli operators.

\textit{Secondly, experimental simulation.}

The details of the experimental simulation will be presented in the next
section. Here we emphasize two problems.

One is how to choose a working initial state for the quantum simulation. In
order to get the spectrum of $H_{p}$, one should select an appropriate
initial state which can lead to a nonzero absorptive spectrum after the
whole procedure of quantum simulation. Note that it is nontrivial to find
such a state. In a recent work\ \cite{wang2}, we proposed a general method
of the selection of an initial state for $N$-qubit quantum simulation of
BCS\ model. This method guarantees the existence of the absorptive peaks in
the $H_{p}$ spectrum and the correspondence between the peaks and the $H_{p}$
enengy levels is clearly known.

The other problem is how to obtain the spectrum of a Hamiltonian by quantum
simulation on an NMR QC. We propose that the simulation scheme has to
include twice Fourier transforms ($ft$s) --- the first $ft$ is from
temporal-domain to frequency-domain of the NMR free induction decay (FID),
and the second $ft $ is of the NMR amplitudes of a peak.

To validate this point, let us briefly review the procedure of the NMR
measure (the first $ft$). NMR internal Hamiltonian in a lab frame is \cite%
{cory2}

\begin{equation}
H_{int}^{lab}=\frac{1}{2}(\sum_{i=1}^{N}\omega _{0}^{i}\sigma
_{z}^{i}+\sum_{j>i=1}^{N}\pi J\sigma _{z}^{i}\sigma _{z}^{j}).
\end{equation}%
NMR measure process under such an internal Hamiltonian is \cite{cory2},

\begin{equation}
S_{NMR}(\omega )\varpropto ft[Tr(e^{-iH_{int}^{lab}t}\rho
_{fin}e^{iH_{int}^{lab}t}\sum_{i=1}^{N}\sigma _{i}^{+})],  \label{nmrmea}
\end{equation}%
where we label the evolution time during the NMR measure as $t$; $%
S_{NMR}(\omega )$ the NMR frequency-domain spectrum; $\rho _{fin}$ the state
before the measurement, \textit{i.e.}, the state after a unitary evolution,
which is, in our case, the evolution with $H_{p}$; $\sigma _{i}^{+}$ the
project operator, $\sigma _{{}}^{+}=\sigma _{x}^{{}}+i\sigma _{y}^{{}}$; $Tr$
the trace operation; $ft$ the Fourier transform applied for the NMR measured
signal which is the function of NMR measure time $t$.

Eq. (\ref{nmrmea}) implies that after the NMR measure, the spectrum is just
that of NMR Hamiltonian self, but not the spectrum of the simulated
Hamiltonian. Thus whatever the initial state or the evolution is, the
spectrum is unchanged after the NMR measurement. From only NMR measure, we
can not get the knowledge of the simulated Hamiltonian. However, the
amplitude in NMR spectrum includes the information of the $H_{p}$ evolution
time. If we select a series of discrete $H_{p}$ evolution time, and measure
the corresponding amplitudes of the each spectrum, then apply another $ft$
on these amplitudes, the result will be the spectrum of the Hamiltonian
simulated. A more formal mathematic derivation for this procedure refers to
\cite{wang2}.

From the above discussion, due to the characteristic of NMR measurement,
twice Fourier transforms are required to get the spectrum of the simulated
Hamiltonian when performing the quantum simulation. The first $ft$ is the
NMR measure applied for the NMR measure time, which obtains the NMR energy
spectrum, and the second $ft$ is applied for the simulated Hamiltonian
evolution time, which obtains the spectrum of $H_{p}$.

\textit{Thirdly, mapping back to the simulated system.}

After the simulation, we are able to get the spectrum of $H_{p}$. However,
we have to remap it back into the BCS physical system and formulate the
energy gap, which is the final aim of quantum simulation. In other words, we
would like to know how the $H_{p}$ spectrum corresponds to the energy gap in
BCS model. The derivation of the relationship between the eigenvalues of $%
H_{p}$ and the energy gap can be found in Ref. \cite{wang2} by the
diagonalization of the spin-analogy Hamiltonian submatrices \cite{wang1} and
numerical calculations \cite{xu}. The energy gap is a function of
eigenvalues corresponding to states $\left\vert 01\right\rangle $ and $%
\left\vert 10\right\rangle $ which span a subspace with one spin-up states,
\textit{i.e.}, one Cooper pair \cite{lawu}. Thus, if we obtain the
eigenvalues of the $H_{p}$ through the quantum simulation, the energy gap in
BCS superconductor can be deduced.

\section{Experimental implementation}

We performed the experiment on a Bruker AVANCE 400 MHz spectrumeter, keeping
the temperature at 300K. The spin system is $^{13}$C-labeled Chloroform
(Cambridge Isotopes) dissolved in 6$d$ acetone. Spins $^{13}C$ and $^{1}H$
are the two qubits and labeled as qubits 1 and 2. The coupling constant is $%
J=214.9$Hz.

Our simulation can be described with 4 stages.

1) Prepare for the working initial state.

The simplest initial state for a two-qubit system is \cite{wang2} $%
\left\vert \psi \right\rangle _{ini}=\left\vert 00\right\rangle +\left\vert
01\right\rangle $. We began with a pseudopure state based on spatial
averaging \cite{cory4,du}, and then apply a Hadamard gate on the second
qubit, say $H_{2}$, to obtain the working initial state,

\begin{equation}
\left\vert 00\right\rangle \overset{H_{2}}{\rightarrow }\left\vert
00\right\rangle +\left\vert 01\right\rangle .  \label{ini}
\end{equation}

2) Realize the unitary transformation.

In this paper, we consider a 2-qubit system and its evolution with $H_{p}$.
The Hamiltonian is

\bigskip
\begin{equation}
H_{p}=\frac{\hbar }{2}[\varepsilon _{1}\sigma _{z}^{1}+\varepsilon
_{2}\sigma _{z}^{2}+V(\sigma _{x}^{1}\sigma _{x}^{2}+\sigma _{y}^{1}\sigma
_{y}^{2})],
\end{equation}%
and the corresponding propagator is

\begin{equation}
U_{p}(\tau )=e^{-iH_{p}\tau /\hbar }=e^{-i\frac{1}{2}[\varepsilon _{1}\sigma
_{z}^{1}+\varepsilon _{2}\sigma _{z}^{2}+V(\sigma _{x}^{1}\sigma
_{x}^{2}+\sigma _{y}^{1}\sigma _{y}^{2})]\tau },  \label{uporigin}
\end{equation}%
where we label the evolution time with $H_{p}$ as $\tau $. Note that we want
to investigate whether the NMR control technique is adapted to be a testbed
for the quantum simulation. In order to check the level of experimental
accuracy to perform this simulation without any error in the unitary
decomposition, we choose a special case of the Hamiltonian (\ref{uporigin})
-- set the parameters $\varepsilon _{1}=\varepsilon _{2}=\varepsilon $, so
as to the four terms in Eq. (\ref{uporigin}) are commuted with each other.
Then Eq. (\ref{uporigin}) can be decomposed exactly into

\begin{equation}
U(\tau )=e^{-i\frac{1}{2}\varepsilon \sigma _{z}^{1}\tau }\cdot e^{-i\frac{1%
}{2}\varepsilon \sigma _{z}^{2}\tau }\cdot e^{-i\frac{1}{2}V\sigma
_{x}^{1}\sigma _{x}^{2}\tau }\cdot e^{-i\frac{1}{2}V\sigma _{y}^{1}\sigma
_{y}^{2}\tau }.  \label{up}
\end{equation}%
Actually, even if $\varepsilon _{1}\neq \varepsilon _{2}$, there is no any
important difference for our simulation procedure.

\begin{figure}[tbp]
\begin{center}
\includegraphics[scale=1]{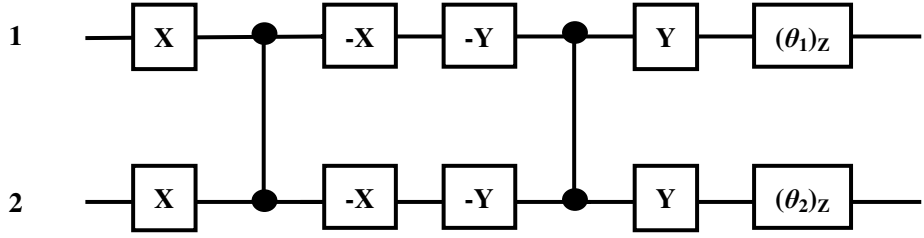}
\end{center}
\caption{Quantum network to realize the evolution of 2-qubit BCS
Hamiltonian. Horizontal lines represent qubits 1 and 2. Time goes from left
to right. The boxes marked by X, -X, Y, -Y denote the $\protect\pi /2$
rotations along the respective directions, and $(\protect\theta _{i})_{z}$ $%
(i=1,2)$ the rotation along z direction with the angle $\protect\theta _{i}$%
. The combination of two dots and a vertical line denotes the J-coupling
evolution,\ \textit{i.e.}, the Ising gate, namely $e^{-i\frac{1}{2}\protect%
\pi J\protect\sigma _{z}^{1}\protect\sigma _{z}^{2}\protect\tau _{3}}$.}
\label{mypic1}
\end{figure}

Experimentally we set both of the spins at their resonance frequency
respectively and use $z$-rotations to realize the first two terms in Eq. (%
\ref{up}), and then, use J-coupling evolutions to realize the last two
terms. With the experimental one-qubit and two-qubit gates, Eq. (\ref{up})
can be expressed into

\begin{align}
U(\tau )& =e^{-i\frac{1}{2}\omega _{1}^{{}}\sigma _{z}^{1}\tau _{1}}\cdot
e^{-i\frac{1}{2}\omega _{2}^{{}}\sigma _{z}^{2}\tau _{2}}  \notag \\
& e^{-i\frac{\pi }{4}\sigma _{y}^{1}}\cdot e^{-i\frac{\pi }{4}\sigma
_{y}^{2}}\cdot e^{-i\frac{1}{2}\pi J\sigma _{z}^{1}\sigma _{z}^{2}\tau
_{3}}\cdot e^{i\frac{\pi }{4}\sigma _{y}^{1}}\cdot e^{i\frac{\pi }{4}\sigma
_{y}^{2}}  \notag \\
& e^{i\frac{\pi }{4}\sigma _{x}^{1}}\cdot e^{i\frac{\pi }{4}\sigma
_{x}^{2}}\cdot e^{-i\frac{1}{2}\pi J\sigma _{z}^{1}\sigma _{z}^{2}\tau
_{3}}\cdot e^{-i\frac{\pi }{4}\sigma _{x}^{1}}\cdot e^{-i\frac{\pi }{4}%
\sigma _{x}^{2}},  \label{u_tau}
\end{align}%
where $\tau _{3}$ is the J-coupling evolution time; $\tau _{1}$, $\tau _{2}$
are the $z$-pulse widths; the r.f. powers $\omega _{1}^{{}}$ and $\omega
_{2}^{{}}$ can be specified by the $90$ degree pulse widths of two spins
respectively \cite{ernst}. The $z$ pulse $e^{-i\frac{1}{2}\sigma
_{z}^{i}\theta _{i}}$ on spin $i$ $(i=1,2)$ with the rotation angle $\theta
_{i}$ is implemented using the combination of $x$ and $y$ pulses $\left(
\frac{\pi }{2}\right) _{x}^{i}-\left( \theta \right) _{y}^{i}-\left( \frac{%
\pi }{2}\right) _{-x}^{i}$ and the rotation angle is $\theta _{i}=\omega
_{i}^{{}}\tau _{i}^{{}}$. The corresponding network to realize $U(\tau )$ is
shown in Fig. \ref{mypic1}.

The parameters in Eq. (\ref{u_tau}) and Eq. (\ref{up}) satisfy

\begin{equation}
\omega _{1}^{{}}\tau _{1}^{{}}=\omega _{2}^{{}}\tau _{2}^{{}}=\varepsilon
\tau ,\text{ }\pi J\tau _{3}=V\tau .  \label{tau2}
\end{equation}

According to the values of the parameters in BCS Hamiltonian (\ref{p}), we
set $\varepsilon _{m}/2\pi \sim 10^{4}$ Hz, $V/2\pi \sim 1$ Hz in the
experimental simulation. For the $H_{p}$ evolution time $\tau $ in Eq. (\ref%
{up}), we select 64 different values --- the initial value is 0, the maximal
value is $\frac{63}{2\pi }$ s, and the increment is $\frac{1}{2\pi }$ s. The
NMR experimental evolution time $\{\tau _{1}$, $\tau _{2}$, $\tau _{3}\}$
can be calculated from the evolution time $\tau $ using the Eq. (\ref{tau2}%
). That means we repeat the experiments for 64 times with the 64 sets of
evolution time $\{\tau _{1}$, $\tau _{2}$, $\tau _{3}\}$. In experiment, the
receiver phases are set with the same phase as that of the measurement of
the first set $\{\tau _{1}$, $\tau _{2}$, $\tau _{3}\}$. We obtain the
relative amplitudes of the finial states by integrating the peaks.

Since the decoherence time \cite{du} ($T_{2}$ relaxation time) of the two
qubits $^{1}$H and $^{13}$C is no more than $3.3$ s and $0.4$ s,
respectively, the whole evolution time should be controlled within the
decoherence time. However, the values of evolution time in the latter sets $%
\{\tau _{1}$, $\tau _{2}$, $\tau _{3}\}$ are so long to exceed the
decoherence time. To solve this problem, we experimentally use a technique
to shorten the evolution time based on the fact that both the $z$-rotation
and the J-coupling evolution are the periodic rotations with the period $%
2\pi $. Therefore, the $z$-pulse width and the J-coupling evolution time
longer than one period can be shortened less than one period.

More specifically, the period of $z$-pulse is $2\pi $ radian, and since we
use the hard pulses in heteronuclear system, the value of one period is $%
\sim 40$ $\mu $s. The period of J-coupling evolution is $2/J$. In our
system, this value is $\sim 10$ ms. Note that compared with the J-coupling
evolution time, the $z$-pulse width is so small and can be ignored. So the
experiment time mainly depends on the J-coupling evolution time. Further
note that there are two segments of the J-coupling evolution in our netwok
(Fig. \ref{mypic1}), the whole time of our experiment is about $20$ ms,
which is well within the decoherence time. In our experiment, the $z$-pulse
width and the J-coupling evolution time are listed in table. \ref{table1}.

\begin{table}[tbp]
\caption{Discrete sets of values of the evolution time used in experiment. $%
\protect\tau _{1}$ is the $z$-pulse width of qubit 1 ($^{13}C$), $\protect%
\tau _{2}$ is the $z$-pulse width of qubit 2 ($^{1}H$), and $\protect\tau %
_{3}$ is the J-coupling evolution\ time. Note the periods of $\protect\tau %
_{1}$, $\protect\tau _{2}$, $\protect\tau _{3}$ are $\sim 40$ $\protect\mu $%
s, $\sim 40$ $\protect\mu $s, $\sim 10$ ms, respectively. When the values
exceed their respective periods, their values can be shortened. See text for
details.}
\label{table1}%
\begin{ruledtabular}
\begin{tabular}{cccc}
&\multicolumn{1}{c}{$\tau_{1}$}&\multicolumn{1}{c}{$\tau_{2}$}&\multicolumn{1}{c}{$\tau_{3}$}\\

\hline

initial values&$0$&$0$&$0$ \\
increment&$17.6$ $\mu$s&$18.0$ $\mu$s&$1.4805$ ms\\
final values&$19.7$ $\mu$s&$20.1$ $\mu$s&$0.2489$ ms\\
\end{tabular}
\end{ruledtabular}
\end{table}

After the evolution, the final state is

\begin{equation}
\left\vert \psi \right\rangle _{fin}=e^{-iH_{p}\tau }\left\vert \psi
\right\rangle _{ini}=e^{-i\varepsilon \tau }\left\vert 00\right\rangle +\cos
(V\tau )\left\vert 01\right\rangle -i\sin (V\tau )\left\vert 10\right\rangle
.  \label{roufin}
\end{equation}%
This state implies that there are two peaks appeared in NMR spectrum -- one
corresponds to the transition $\left\vert 00\right\rangle \leftrightarrow
\left\vert 10\right\rangle $, and the other corresponds to $\left\vert
00\right\rangle \leftrightarrow \left\vert 01\right\rangle $ depended on the
nonvanishing coefficients.

\begin{figure}[tbp]
\begin{center}
\includegraphics[scale=0.6]{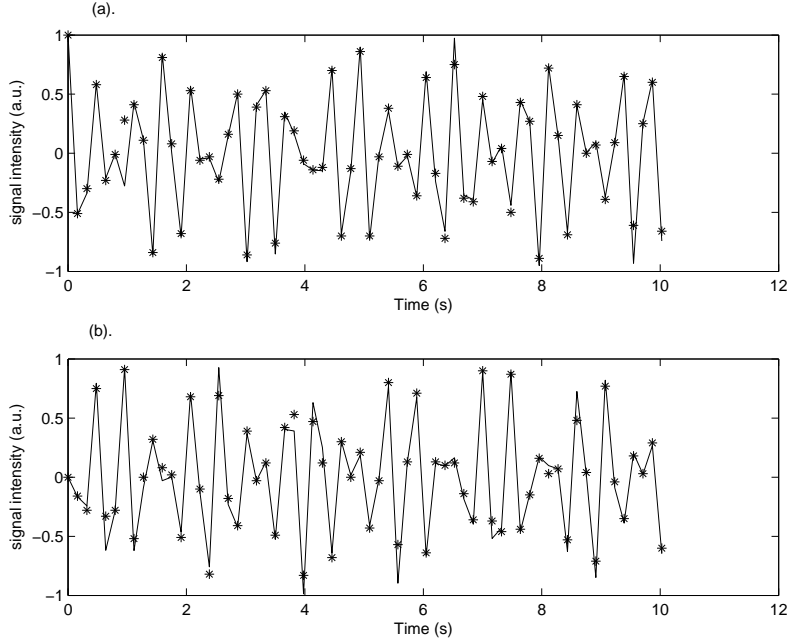}
\end{center}
\caption{The real (a) and the imaginary parts (b) of the NMR measured
amplitudes. The solid lines are fitted to the theoretical expectation. The
stars are the experimental data. Note the horizontal axes is the evolution
time $\protect\tau $, from $0$ to $63/2\protect\pi $ s.}
\label{mypic2}
\end{figure}

3) The 1st $ft$. Since NMR spectrum of the eigenvalues is unchanged, the
positions of the peaks include no useful information. What we need is to
record the amplitudes of the peaks. For the initial state we select, there
are two peaks in the final spectrum, as shown in Eq. (\ref{roufin}) --- one
is the peak of $^{13}C$, another is the peak of $^{1}H$. We measure the
peaks of $^{1}$H, and record the real and imagery parts of the amplitudes
corresponding to the different evolution time through the integration of the
peaks. By substituting Eq. (\ref{roufin}) into Eq. (\ref{nmrmea}), the
amplitudes of the peak of $^{13}C$ and $^{1}H$ theoretically satisfied the
equations

\begin{eqnarray}
Amp_{1} &=&i\sin (V\tau )[\cos (\varepsilon \tau )+i\sin (\varepsilon \tau
)],  \label{A1} \\
Amp_{2} &=&\cos (V\tau )[\cos (\varepsilon \tau )+i\sin (\varepsilon \tau )],
\label{A2}
\end{eqnarray}%
respectively. The measured amplitudes fit the Eq. (\ref{A2}). The real and
imaginary parts of the experimentally measured amplitudes are shown in Fig. %
\ref{mypic2}.

\begin{figure}[tbp]
\begin{center}
\includegraphics[scale=0.6]{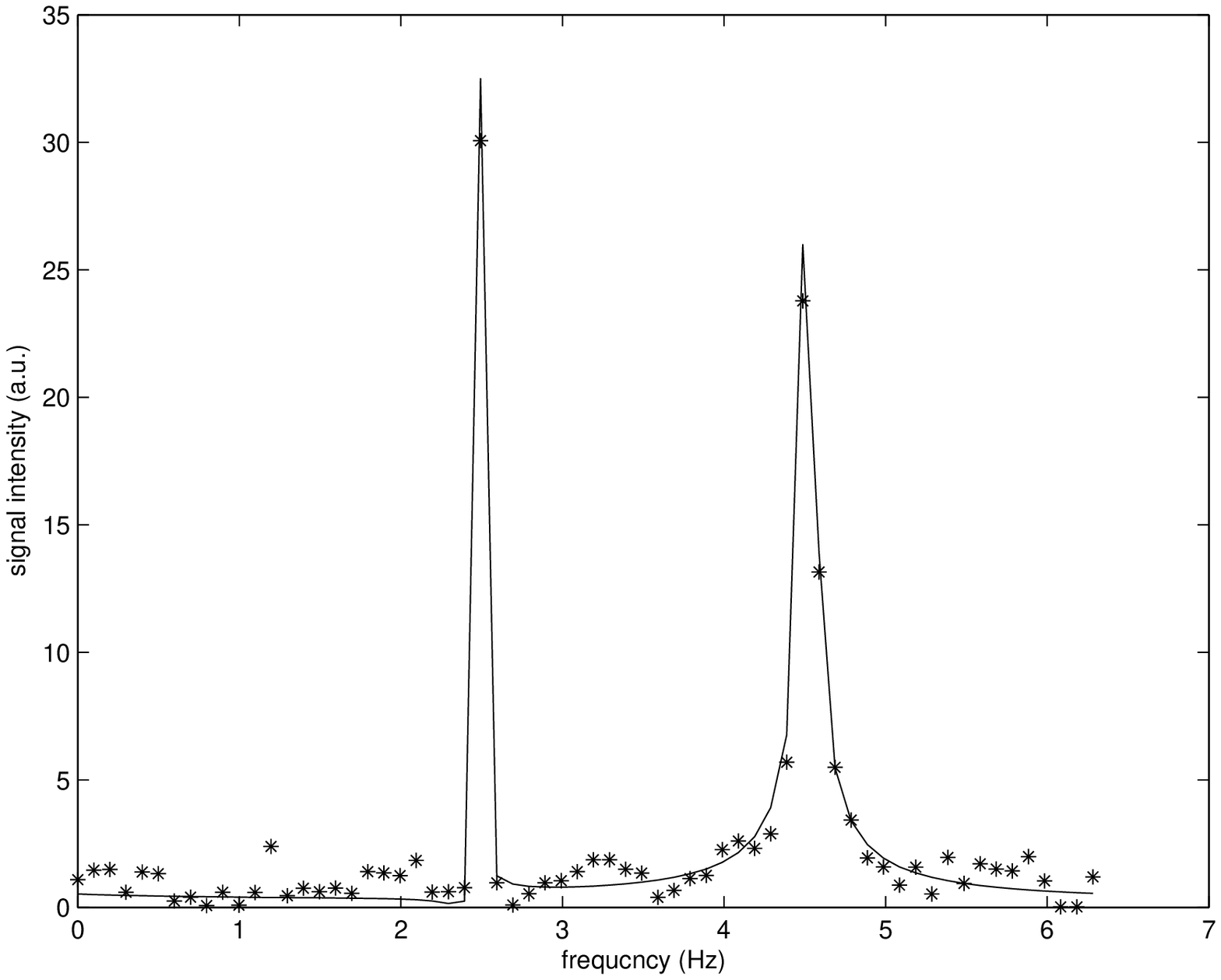}
\end{center}
\caption{The spectrum of the Hamiltonian $H_{p}$ obtained by the second
Fourier transform. The solid line is fitted to the theoretical expectation.
The distance between two peaks shows the difference of the eigenvalues of $%
H_{p}$ is $2$ Hz, which is same as the theoretical calculation.}
\label{mypic3}
\end{figure}

4) The 2nd $ft$. Apply a discrete $ft$ on the amplitudes, and obtain the
spectrum of $H_{p}$, shown in Fig. \ref{mypic3}. The distance of the two
peaks is $2$ Hz from the figure. The value is same as the theoretical result
which can trivially diagonalize the Hamiltonian (\ref{p}) and calculate the
energy difference corresponding to the eigenstate $\left\{ \left\vert
01\right\rangle ,\left\vert 10\right\rangle \right\} $. The difference of
the eigenvalues is just the information we want to know from the Hamiltonian
$H_{p}$. Since $H_{p}$ is the mapped one from the BCS Hamiltonian, we can
remap the eigenvalues of $H_{p}$ to get the energy gap in BCS Hamiltonian
\cite{wang2}. The errors mainly arise as a result of the imperfect pulses
and delays as well as the variability of the measurement.

\section{Conclusion}

In this paper, a concrete experimental scheme of quantum simulation on NMR
QC is presented. Two important components in our scheme are the selection of
an appropriate working initial state and the usage of twice Fourier
transforms. Based on this scheme, we simulate the evolution of the BCS
Hamiltonian with an experimental technique to reduce the system evolution
within the decoherence time. The simulation results show the BCS energy
spectrum agrees with the theoretical value. This provides a practical method
to obtain the characteristic of the complex physical system, and shows the
potential power of the QC in quantum simulation.

It is believed that quantum simulation is possible to solve the quantum
dynamics. The ability to perform an efficient quantum simulation requires
many degrees of freedom (perhaps 10$\sim $100 qubits in the next generation
QC \cite{lawu}). In this paper, we are putting our research on those issues
associated with problems that are difficult for quantum many-body theory to
solve on classical computer. Although a 2-qubit quantum simulation of BCS
model is shown in experiment, the fact that even small numbers of qubits
leads to new insight into the mechanism of QC how to perform quantum
simulation, and permit accurate prediction of the dynamics of complex
quantum system if the technique developed future.

Clearly, a number of challenges for the efficient simulation of physical
system still remain at present \cite{ortiz}: Can we find the physical
results in the simulated system after the algebras remapping? How can we
realize a complex network of multi-qubit system with the enough accuracy?
Can we find the a physical simulation problem that a QC can solve but not
the classical computer today? Nevertheless, we are sure that the concept
presented here establishes a very first step to simulate a real physical
system on QC.

\bigskip

\begin{acknowledgments}
We thank Xiao-San Ma, Hao You, Wan-Qing Niu, and Ping Zou for discussions.
X.-D Y thanks Jun Luo for continuous encouragement. This project was
supported by the National Fundamental Research Program (Grant No.
2001CB309309 and No. 2001CB309310), the NSFC (Grant No. 60173047 and No.
10075041), China Post-doctoral Science Foundation, and the Natural Science
Foundation of Anhui Province. We also thank supports from the ASTAR (Grant
No. 012-104-305).
\end{acknowledgments}

\bigskip \textit{Note added. --} After this work was completed, we became
aware of a similar work appeared in quant-ph/0410106 by C. Negrevergne, R.
Somma, G. Ortiz, E. Knill, and R. Laflamme, under the title \textit{liquid
state NMR simulation of quantum many-body problem}.

\end{document}